\documentclass[prd,preprint,floatfix,nofootinbib,amsmath]{revtex4}
\usepackage{rotate,subfig}
\usepackage{epsfig}
\usepackage{graphicx}
\usepackage{pstricks}

\catcode`\@=11
\def\sla@#1#2#3#4#5{{%
 \setbox\z@\hbox{$\m@th#4#5$}%
 \setbox\tw@\hbox{$\m@th#4#1$}%
 \dimen4\wd\ifdim\wd\z@<\wd\tw@\tw@\else\z@\fi
 \dimen@\ht\tw@
 \advance\dimen@-\dp\tw@ \advance\dimen@-\ht\z@
 \advance\dimen@\dp\z@
 \divide\dimen@\tw@ \advance\dimen@-#3\ht\tw@
 \advance\dimen@-#3\dp\tw@ \dimen@ii#2\wd\z@
 \raise-\dimen@\hbox to\dimen4{%
 \hss\kern\dimen@ii\box\tw@\kern-\dimen@ii\hss}%
 \llap{\hbox to\dimen4{\hss\box\z@\hss}}}}

\def\slashed#1{%
 \expandafter\ifx\csname sla@\string#1\endcsname\relax
{\mathpalette{\sla@/00}{#1}}
\fi}
\def\declareslashed#1#2#3#4#5{%
 \expandafter\def\csname sla@\string#5\endcsname{%
#1{\mathpalette{\sla@{#2}{#3}{#4}}{#5}}}}
 \catcode`\@=12
\declareslashed{}{/}{.08}{0}{D}
 \declareslashed{}{/}{.1}{0}{A}
 \declareslashed{}{/}{0}{-.05}{k}
 \declareslashed{}{/}{.1}{0}{\partial}
 \declareslashed{}{\not}{-.6}{0}{f}
\def\lsim{\mathrel {\vcenter {\baselineskip 0pt \kern 0pt
    \hbox{$<$} \kern 0pt \hbox{$\sim$} }}}
\def\gsim{\mathrel {\vcenter {\baselineskip 0pt \kern 0pt
    \hbox{$>$} \kern 0pt \hbox{$\sim$} }}}
\def\slashchar#1{\setbox0=\hbox{$#1$}           
 \dimen0=\wd0                                 
  \setbox1=\hbox{/} \dimen1=\wd1               
\ifdim\dimen0>\dimen1                        
  \rlap{\hbox to \dimen0{\hfil/\hfil}}      
  #1                                        
  \else                                        
 \rlap{\hbox to \dimen1{\hfil$#1$\hfil}}   
   /                                         
  \fi}                                         %
\def\cpto{\mathrel {\vcenter {\baselineskip 0pt \kern 0pt
    \hbox{$CP$} \kern 0pt \hbox{$\longrightarrow$} }}}
\def\cptof{\mathrel {\vcenter {\baselineskip 0pt \kern 0pt
    \hbox{$~CP$} \kern 0pt \hbox{$\longleftrightarrow$} }}}

\begin{document}

\baselineskip=15pt
\preprint{}

\title{$CP$-odd correlations using jet momenta from $t\bar{t}$ events at the Tevatron}

\author{Sudhir Kumar Gupta and G. Valencia}

\email{skgupta@iastate.edu,valencia@iastate.edu}

\affiliation{Department of Physics, Iowa State University, Ames, IA 50011.}

\date{\today}

\vskip 1cm
\begin{abstract}

We discuss $T$-odd correlations between jet and lepton momenta in $t\bar{t}$ events at the Tevatron that can be used to search for $CP$ violation. We identify correlations suitable for the lepton plus jets and purely hadronic top-quark pair decay channels. As an example of $CP$ violation we consider the top-quark  anomalous couplings, including its chromo-electric dipole moment,  and we estimate the limits that can be placed at the Tevatron. 

\end{abstract}


\maketitle

\section{Introduction}

Looking for new sources of $CP$ violation remains one of the important goals of high energy colliders. Processes that have been considered before include production and decay of top quark pairs\cite{reviews,Donoghue:1987ax,ttpairs,eeanom,Antipin:2008zx,Gupta:2009wu},  production and decay of electroweak gauge bosons  \cite{cpgaugeb}, and production and decay of new particles \cite{newpart}. 

The Tevatron has now observed hundreds of $t\bar{t}$ events and will eventually collect a few thousand. With this in mind, it is interesting to revisit the issue of possible $CP$ violation in these events. Under optimal conditions, a sample of a few thousand events would have a statistical sensitivity to a $CP$ violating asymmetry at the few percent level. 

In this paper we study $T$-odd triple product correlations of the sort first discussed in Ref.~\cite{Donoghue:1987ax,tripprods} and find observables suitable for $t\bar t$ events in which the top quark pairs decay into a lepton plus jets or purely hadronically. Of particular interest are observables that only require reconstruction of two $b$ jets (but there is no need to distinguish between $b$ and $\bar b$); one or no hard leptons; and non-$b$ jets ordered by $p_T$.

It is well known that $CP$ violation in the standard model (SM) is too small to induce a signal at an observable level in the processes we consider.  We will discuss two $CP$ violating effective interactions that should serve as benchmarks for the sensitivity of the Tevatron to  $CP$ violation in $t\bar{t}$ events. 

As in our LHC study \cite{Gupta:2009wu},  we parametrize $CP$ violation using anomalous top-quark couplings.  The $t\bar{t}$ production process is modified relative to the SM by the chromo-electric dipole moment (CEDM) of the top-quark via the interaction
\begin{eqnarray}
{\cal L}_{cdm}&=&-ig_s\frac{\tilde{d}}{2}\bar{t}\, \sigma_{\mu\nu}\gamma_5  \, G^{\mu\nu}\, t,
\label{dtilde}
\end{eqnarray}
where $g_s$ is the strong coupling constant and $G^{\mu\nu}$ is the usual gluon field strength tensor.  The CEDM is induced in principle by any theory that violates $CP$, and estimates for its size in several models can be found in Ref.~\cite{reviews}.  Typical estimates presented in Ref.~\cite{reviews} for the size of  $\tilde{d}$ suggest that it may be too small to yield observable signals at the Tevatron. Nevertheless, we view the study of this coupling at the Tevatron as a valuable preliminary to future LHC studies. 

We also consider $CP$ violation in the decay vertices $t\to b W^+$ and $\bar t\to \bar b W^-$ via the anomalous coupling $f$ defined by\footnote{As discussed in the literature, other anomalous couplings will not interfere with the SM and we will not consider them here ~\cite{delAguila:2002nf}.},
\begin{eqnarray}
\Gamma^\mu_{Wtb} &=& 
-\frac{g}{\sqrt{2}} \, V_{tb}^\star \,\bar{u}(p_b) \left[ \gamma_\mu P_L -i \tilde{f} e^{i(\phi_f+\delta_f)} \sigma^{\mu\nu} (p_t-p_b)_\nu  P_R \right] u(p_t) \nonumber \\
\nonumber \\
\bar\Gamma^\mu_{Wtb} &=&
-\frac{g}{\sqrt{2}} \, V_{tb} \,\bar{v}(p_{\bar{t}}) \left[ \gamma_\mu P_L-
i \tilde{f} e^{i(-\phi_f+\delta_f)} \sigma^{\mu\nu} (p_{\bar t}-p_{\bar b})_\nu P_L \right] v(p_{\bar b}),
\label{ftilde}
\end{eqnarray}
In Eq.~\ref{ftilde} we have explicitly split the phase of $\tilde{f}$ into a $CP$ violating phase $\phi_f$ and a $CP$ conserving, unitarity, phase $\delta_f$. 

The $CP$ violating anomalous couplings, $\tilde{d}$ and $\tilde{f}\sin\phi_f$,  have been recently revisited vis-a-vis the upcoming LHC experiments in Ref.~\cite{Antipin:2008zx} and Ref.~\cite{Gupta:2009wu}. In Ref.~\cite{Antipin:2008zx}, general results were derived for the $T$-odd correlations induced by these two couplings for both gluon fusion and light $q\bar{q}$ annihilation $t\bar{t}$ production processes. In the appendix we specialize those general results to the specific processes that are relevant for the Tevatron. In Ref.~\cite{Gupta:2009wu} a numerical analysis was carried out for LHC concentrating on the dilepton signal, which is not viable at the Tevatron due to the small number of events. The new signals we discuss in this paper pertain to the lepton plus jets and all hadronic decay modes of the top-quark pairs and can also be used at LHC.

\section{Observables}

In Ref.~\cite{Antipin:2008zx} all the linearly independent $T$-odd  correlations induced by anomalous top-quark couplings were identified. From these we need to project out the ones that are most suitable for the Tevatron  and two considerations come into play. The first one, already discussed in our application to the LHC in Ref.~\cite{Gupta:2009wu}, is that we want to use only momenta that can be reconstructed experimentally. The second one is that, due to the low statistics at the Tevatron, we will be dealing with at least one hadronic decay of the $W$ boson. 

We will consider the following correlations
\footnote{Here we use the Levi-Civita tensor contracted with four vectors $\epsilon(a,b,c,d) \equiv \epsilon_{\mu \nu \alpha \beta} a^\mu b^\nu c^\alpha d^\beta$ with the sign convention $\epsilon_{0123}=1$. We also use $s,t,u$ to refer to the parton level Mandelstam variables for $q\bar{q} \to t \bar{t}$.}:

\begin{itemize}

\item For the lepton (muon) plus jets process $p\bar{p} \to t\bar{t}\to b\bar{b} \mu j_1 j_2 + \slashed{E}_T$:
\begin{eqnarray}
{\cal O}_1 &=& \epsilon(p_t,p_{\bar{t}},p_b,p_{\bar b}) \,\,
\xrightarrow[]{t\bar t ~CM}\,\, \propto \,\, \vec{p}_t\cdot (\vec{p}_b \times \vec{p}_{\bar b})
\nonumber \\
{\cal O}_2 &=& \epsilon(P,p_b+p_{\bar b},p_\ell,p_{j1}) \,\,
\xrightarrow[]{lab}\,\, \propto \,\, (\vec{p}_b +\vec{p}_{\bar b})\cdot (\vec{p}_\ell \times \vec{p}_{j1})
\nonumber \\
{\cal O}_3 &=&  Q_\ell \, \epsilon(p_b,p_{\bar b},p_\ell,p_{j1})  \,\,
\xrightarrow[]{b\bar b ~CM}\,\, \propto \,\, Q_\ell\,\vec{p}_b \cdot (\vec{p}_\ell \times \vec{p}_{j1})
\nonumber \\
{\cal O}_4 &=&  Q_\ell \, \epsilon(P,p_b-p_{\bar b},p_\ell,p_{j1}) \,\,
\xrightarrow[]{lab}\,\, \propto \,\, Q_\ell\,(\vec{p}_b -\vec{p}_{\bar b})\cdot (\vec{p}_\ell \times \vec{p}_{j1})
\nonumber \\
{\cal O}_7 &=&  \tilde{q}\cdot(p_b-p_{\bar b})\, \epsilon(P,\tilde{q},p_b,p_{\bar b}) \,\,
\xrightarrow[]{lab}\,\, \propto \,\, \vec{p}_{beam}\cdot (\vec{p}_b -\vec{p}_{\bar b})  \,\vec{p}_{beam}\cdot (\vec{p}_b\times\vec{p}_{\bar b}).
\label{prodco1}
\end{eqnarray}

\item For the multi-jet process $p\bar{p} \to t\bar{t}\to b\bar{b} j_1j_2j_{1'}j_{2'}$:
\begin{eqnarray}
{\cal O}_1 &=& \epsilon(p_t,p_{\bar{t}},p_b,p_{\bar b}) \,\,
\xrightarrow[]{t\bar t ~CM}\,\, \propto \,\, \vec{p}_t\cdot (\vec{p}_b \times \vec{p}_{\bar b})
\nonumber \\
{\cal O}_5 &=& \epsilon(p_b,p_{\bar b},p_{j1},p_{j1'})  \,\,
\xrightarrow[]{b\bar b ~CM}\,\, \propto \,\, \vec{p}_b \cdot (\vec{p}_{j1} \times \vec{p}_{j1'})
\nonumber \\
{\cal O}_6 &=&  \epsilon(p_b,p_{\bar b},p_{j1}+p_{j2},p_{j1'}+p_{j2'})\,\,
\xrightarrow[]{t\bar t ~CM}\,\, \propto \,\, (\vec{p}_{j1}+\vec{p}_{j2})\cdot (\vec{p}_b \times \vec{p}_{\bar b})
\nonumber \\
{\cal O}_7 &=&  \tilde{q}\cdot(p_b-p_{\bar b})\, \epsilon(P,\tilde{q},p_b,p_{\bar b}) \,\,
\xrightarrow[]{lab}\,\, \propto \,\, \vec{p}_{beam}\cdot (\vec{p}_b -\vec{p}_{\bar b})  \,\vec{p}_{beam}\cdot (\vec{p}_b\times\vec{p}_{\bar b}).
\label{prodco2}
\end{eqnarray}

\end{itemize}

In Eqs.~\ref{prodco1}~and~\ref{prodco2} we have shown two expressions for each of the correlations. The first one is valid in any frame and in particular can be used in the lab frame. The second one shows the correlation in a particular frame in which it reduces to a simple triple vector product. 
In these expressions $P$ is the sum of the proton and antiproton four-momenta; $\tilde{q}$ is the difference of the proton and antiproton four-momenta; $p_{b,{\bar b}}$ refers to the $b$ or $\bar b$ jet momenta; $p_\ell$ refers to the momenta of a lepton that has been identified as originating from $t$ or $\bar t$ decay in lepton plus jets events (in our analysis we only consider muons);  $p_{j1},~p_{j2}$ refer to {\bf non}-$b$ jets ordered by $p_T$ (hardness) that reconstruct a $W$; primes denote the two jets associated with the second $W$ in the all hadronic case.  In ${\cal O}_{5,6}$ it is not necessary to distinguish $b$ and 
$\bar b$ jets. It is only necessary to associate $j_{1,2}$ with one of the $b$ jets and $j^\prime_{1,2}$ with the other one when reconstructing the top-quark pair event. The hardness of the jet is defined in the usual way, the hardest jet being that with the largest transverse momentum, i.e. $p_{T_{j_1}} > p_{T_{j_2}}$, and $Q_\ell$ is the lepton charge (for some of the monoleptonic signals, lepton charge id is needed).

Notice that some of the correlations require differentiating between the $b$ and $\bar{b}$ jets but others don't. In addition, ${\cal O}_1$ requires the reconstruction of the top momenta. This correlation is the one closest to the form that appears in the parton level calculation, and in a perfect reconstruction situation it is identical to ${\cal O}_6$. In addition, two of the examples given, ${\cal O}_1$  and ${\cal O}_7$ can be used for both processes. Finally we note that there are many other possibilities that we have not listed.

All the correlations listed above are $CP$ odd, as can be seen most easily in the specific reference frames given. For example, ${\cal O}_2$ in the lab frame becomes
\begin{eqnarray} 
{\cal O}_2 & \xrightarrow[]{lab} & \sqrt{S}\left[(\vec p_b+\vec p_{\bar b})\cdot \left((\vec p_{\mu^+} \times \vec p_{j1})+(\vec p_{\mu^-} \times \vec p_{\bar{j}1})\right)\right] \nonumber \\
& \xrightarrow[]{CP} & (-)\sqrt{S}\left[(\vec p_b+\vec p_{\bar b})\cdot \left((\vec p_{\mu^-} \times \vec p_{\bar{j}1})+(\vec p_{\mu^+} \times \vec p_{j1})\right)\right].
\label{example}
\end{eqnarray}
Eq.~\ref{example} also clarifies what is meant by ${\cal O}_2$: events with a $\mu^+$ and a $W^-$ decaying to two jets will contribute to the first term in the sum in the first line. Events with a $\mu^-$ and a $W^+$ decaying to two jets contribute to the second term. The assignment $\vec p_{\bar{j}1} \xrightarrow{CP} - \vec p_{j1}$ on the second line of Eq.~\ref{example} states that if $CP$ is conserved, the probability for a given jet originating from a quark $q$ in a two jet $W^+$ decay to be the hardest one, is equal to the probability for the corresponding jet originating from the anti-quark $\bar{q}$ in a two jet $W^-$ decay to be the hardest one. These statements are verified in our numerical simulations both explicitly and by the fact that the asymmetry is induced by $CP$ violating couplings but vanishes for $CP$ conserving ones. In an experimental analysis it will be important to implement additional cuts in a way that is $CP$ blind, typically requiring the same cuts for particles and anti-particles.

Use of the lepton charge in some of the correlations allows us to construct  $CP$ odd and $CP$ even correlations with the same set of momenta. We exploit this to construct the $T$-odd (but $CP$ even) correlations sensitive to strong phases:
\begin{eqnarray}
{\cal O}_a &=&  \epsilon(P,p_b-p_{\bar b},p_\ell,p_{j1}) \,\,
\xrightarrow[]{lab}\,\, \propto \,\, (\vec{p}_b -\vec{p}_{\bar b})\cdot (\vec{p}_\ell \times \vec{p}_{j1})
\nonumber \\
{\cal O}_b &=& Q_\ell \, \epsilon(P,p_b+p_{\bar b},p_\ell,p_{j1})  \,\,
\xrightarrow[]{lab}\,\, \propto \,\, Q_\ell\,(\vec{p}_b +\vec{p}_{\bar b})\cdot (\vec{p}_\ell \times \vec{p}_{j1})
\nonumber \\
{\cal O}_c &=&   \epsilon(P,p_b+p_{\bar b},p_{j1},p_{j1'}) \,\,
\xrightarrow[]{lab}\,\, \propto \,\, (\vec{p}_b +\vec{p}_{\bar b})\cdot (\vec{p}_{j1} \times \vec{p}_{j1'})
.
\label{prodco3}
\end{eqnarray}
The first two have $CP$-odd analogues in ${\cal O}_4$ and ${\cal O}_2,$ respectively.

Our observables will be  the lab frame distributions $d\sigma/d{\cal O}_i$ for the correlations listed above, as well as their associated integrated counting asymmetries
\begin{eqnarray}
A_i &\equiv & \frac{N_{events}({\cal O}_i >0)-N_{events}({\cal O}_i<0)}{N_{events}({\cal O}_i>0)+N_{events}({\cal O}_i<0)},
\label{asym}
\end{eqnarray}
the denominator being just the total number of events in all cases. 
When our numerical results for the integrated asymmetries are very small we distinguish between very small asymmetries and vanishing asymmetries as described in Ref.~\cite{Gupta:2009wu}.

\section{Numerical Analysis}

Our numerical study in this paper corresponds to the implementation of analytic results presented in Ref.~\cite{Antipin:2008zx}. The $T$-odd correlations for the parton level processes that are relevant at the Tevatron are not explicitly written in Ref.~\cite{Antipin:2008zx}, so we present them in the Appendix for convenience. 
The numerical studies are performed with the aid of Madgraph \cite{Stelzer:1994ta,Alwall:2007st,Alwall:2008pm} following the procedure outlined in Ref.~\cite{Gupta:2009wu}. For the lepton plus jets channel, we begin 
with the standard model processes  $q{\bar q}~ ({\rm or~}gg)\to t{\bar t} \to b \ell^+ \nu {\bar b} {\bar u} d~ ({\rm or~} {\bar b} \ell^- {\bar \nu} {b} {\bar d} u)$  implemented in
Madgraph according to the decay chain feature described in Ref.~\cite{Alwall:2008pm}. This decay chain feature is chosen for consistency with the approximations in the analytical calculation of the $CP$ violating interference term presented in Ref.~\cite{Antipin:2008zx}, in which the narrow width approximation is used for the intermediate top quark and $W$ boson states. The expressions from Ref.~\cite{Antipin:2008zx} (Eqs.~\ref{formfactors}-\ref{cpdecsqq}) are then added to the spin and color averaged matrix element squared for the SM (which Madgraph calculates automatically) and the resulting code is used to generate events. A similar procedure is followed for the purely hadronic decay of $t\bar t$ with the relevant parton level processes. In this case both $W$'s decay into a pair of quarks and we only consider the final states $u,d,s,c$ without Cabibbo mixing. 
The code used to generate events is, therefore, missing the terms that are completely due to new physics: those proportional to the anomalous couplings squared. This approximation is justified because those terms do not generate $T$-odd correlations. In addition, as long as the conditions that allow us to write the new physics in terms of anomalous couplings remain valid, their contribution to the total cross-section is small.

For event generation we  require the top quark and $W$ boson intermediate states to be within 15 widths of their mass shell, and two sets of cuts. The first set of cuts includes a minimum transverse momentum for all leptons and jets, a minimal separation between them, and a pseudorapidity acceptance range:
\begin{equation}             
p_{T_{\mu, j}} > 20~{\rm GeV},\,\, p_{T_{b,{\bar b}}} > 25 ~{\rm GeV}, \,\,
|\eta_i| < 2.5, \,\, \Delta R_ {ik} = \sqrt{{(\eta_i - \eta_k)}^2 + {(\phi_i - \phi_k)}^2}> 0.4
\label{cutone}
\end{equation}     
with $i,k = b, {\bar b}, j, \mu$. 

For the second set of cuts (in the lepton plus jets channel) we add a missing transverse energy requirement
\begin{equation}    
\slashed{E}_T > 30~{\rm GeV}.
\label{twocut} 
\end{equation}    
We use SM parameter values as in Madgraph, except for $m_b=0$; and we use the CTEQ-6L1 parton distribution functions.

\subsection{Process $p\bar{p}  \to t \bar{t} \to b \bar{b} \mu j_1 j_2 + \slashed{E}_T $}

We first estimate the counting asymmetries  by generating $10^6$ events for each of the four cases: $\tilde{d} = 5 \times 10^{-3}~{\rm GeV}^{-1}$; $\tilde{f}\,\sin\phi_f= 5 \times 10^{-3}~{\rm GeV}^{-1}$; $\tilde{f}\,\sin\delta_f= 5 \times 10^{-3}~{\rm GeV}^{-1}$ and $\tilde{d} =\tilde{ f} =0$. These cases correspond to $CP$ violation in the production vertex, $CP$ violation in the decay vertex, strong phases in the decay vertex and the lowest order SM respectively. The relatively large number $5 \times 10^{-3}~{\rm GeV}^{-1}$ is chosen to facilitate distinguishing signals from statistical fluctuations. Once we establish a non-zero asymmetry we can cast our result as a function of the anomalous couplings since the asymmetries are linear in them. 
As mentioned above, we include the new physics only through its interference with the SM. Since these $T$-odd terms are also $P$ odd, they do not affect the cross-sections as they integrate to zero. For this reason the total number of events is the same as in the standard model.  

The results with the set of cuts Eq.~\ref{cutone} are shown in Table~\ref{t:basic1}. After these cuts are applied there remain approximately $6.7 \times 10^5$ events, leading to the $3\sigma$ statistical sensitivity shown in the first column. The results show that all the $CP$-odd correlations vanish for the two $CP$ conserving cases (SM and $\tilde{f}\,\sin\delta_f$), and that the $CP$-even correlations vanish except for the $CP$ conserving case with unitarity phases, $\tilde{f}\,\sin\delta_f$. This establishes numerically that, at least at this level of sensitivity, there is no $CP$ conserving contamination of the $CP$-odd signals or vice versa.

\begin{table}[h]
\centering
\begin{tabular}{| c | c | c |c|c|c|c|c|c|}
\hline\hline
 & $3/\sqrt{N}$&$A_1$&$A_2$&$A_3$&$A_4$&$A_7$&$A_a$&$A_b$\\
\hline
${\tilde d}$&3.7&-66.9&-37.4&-100.6&75.8&40.4&-3.4&-1.8\\
${\tilde f}\sin\phi_f$&3.7&-7.2&-60.8&-8.2&-36.7&10.6&0.6&-1.9\\
${\tilde f}\sin\delta_f$&3.7&-0.8&-1.0&0&-1.7&1.9&-49.3&-51.9\\
\hline
SM&3.7&2.6&-0.6&0.4&0.1&0.2&-0.4&0.3\\
\hline\hline
\end{tabular}\caption{Integrated asymmetries with cuts given in Eq.~\ref{cutone} for  $\tilde{d}$,  ${\tilde f}\sin(\phi_f,\delta_f)$ $= 5 \times 10^{-3}~{\rm GeV}^{-1}$ in units of $10^{-3}$, and the SM. }
\label{t:basic1}
\end{table}

In Table~\ref{t:final1} we show the same results with the additional missing $E_T$ requirement of Eq.~\ref{twocut}. This additional cut further reduces the number of generated events to about $6\times 10^5$, and slightly decreases the statistical sensitivity. The effect of this cut is minimal on all asymmetries, making it very desirable for reducing background. 

\begin{table}[h]
\centering
\begin{tabular}{| c | c | c |c|c|c|c|c|c|}
\hline\hline
 & $3/\sqrt{N}$&$A_1$&$A_2$&$A_3$&$A_4$&$A_7$&$A_a$&$A_b$\\
\hline
${\tilde d}$&3.9&-66.4&-38.9&-102.3&76.5&36.4&-3.0&-1.4\\
${\tilde f}\sin\phi_f$&3.9&-17.2&-66.8&-18.8&-30.8&7.0&0.7&-2.3\\
${\tilde f}\sin\delta_f$&3.9&0.4&-1.1&1.6&-3.1&1.0&-44.1&-56.8\\
\hline
SM&3.9&2.5&-0.5&0.5&-0.2&0.4&-0.3&0.6\\
\hline\hline
\end{tabular}
\caption{Integrated asymmetries  with cuts given in Eqs.~\ref{cutone},~\ref{twocut} for  $\tilde{d}$,  ${\tilde f}\sin(\phi_f,\delta_f)$ $= 5 \times 10^{-3}~{\rm GeV}^{-1}$ in units of $10^{-3}$, and the SM. }
\label{t:final1}
\end{table}

Now we summarize our results for the asymmetries in the process $p\bar{p}  \to t \bar{t} \to b \bar{b} \mu j_1 j_2 + \slashed{E}_T $ with cuts given in Eqs.~\ref{cutone},~\ref{twocut} in terms of the dimensionless anomalous couplings
\begin{eqnarray}
d_t \equiv  \tilde{d} \, m_t, && 
f_t \equiv  \tilde{f}\, m_t
\label{dimcoup}
\end{eqnarray}
with $m_t = 171.2$~GeV. We find, 
\begin{eqnarray}
A_1 &=& -0.078\, d_t -0.020\, f_t\sin\phi_f \nonumber \\
A_2 &=& -0.045\, d_t -0.078\, f_t\sin\phi_f \nonumber \\
A_3 &=& -0.12\, d_t -0.022\, f_t\sin\phi_f \nonumber \\
A_4 &=& 0.089\, d_t -0.036\, f_t\sin\phi_f \nonumber \\
A_7 &=& 0.043\, d_t +0.008\, f_t\sin\phi_f \nonumber \\
A_a &=& -0.052\, f_t\sin\delta_f \nonumber \\
A_b &=& -0.066\, f_t\sin\delta_f .
\label{notoprec1}
\end{eqnarray}

\begin{figure}
\centering
\includegraphics[angle=0, width=1\textwidth]{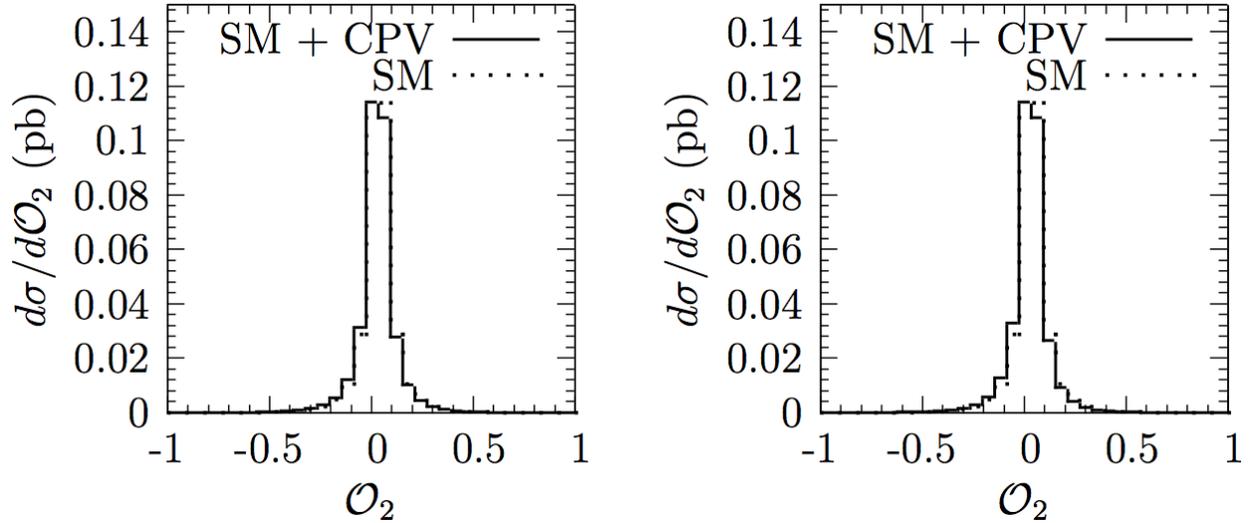}

\caption{Differential distributions $d\sigma/d{\cal O}_2$ for (a) $\tilde{d}=5\times 10^{-3}~{\rm GeV}^{-1}$ and (b) $\tilde{f}\sin\phi_f=5\times 10^{-3}~{\rm GeV}^{-1}$ compared to the SM. ${\cal O}_2$ is given in units of $m_t^4$.}
\label{f:fig2}
\end{figure}

In addition to the integrated counting asymmetries one can look for asymmetries in the distributions $d\sigma/d{\cal O}_i$. In Figure~\ref{f:fig2} we compare the distributions for $d\sigma/d{\cal O}_2$ induced by (a) $\tilde{d}=5\times 10^{-3}~{\rm GeV}^{-1}$ and (b) $\tilde{f}\sin\phi_f=5\times 10^{-3}~{\rm GeV}^{-1}$ to the SM.

\begin{figure}
\centering
\includegraphics[angle=-90, width=0.7\textwidth]{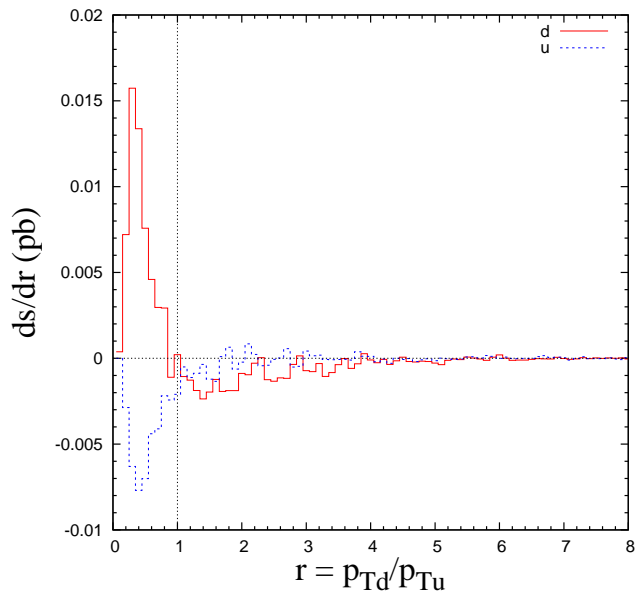}
\caption{Differential distributions of $s$, the numerator of $\tilde{A}_{2d}$ (or $\tilde{A}_{2u}$) ,  with respect to $r$, the ratio of $d$-quark transverse momentum to $u$-quark transverse momentum in $t\to b u \bar{d}$ or $\bar t \to \bar b \bar u d$ decay.}
\label{f:fig1}
\end{figure}

It is instructive to discuss $A_2$ in some detail to understand the role of the hardest jet momenta. The lepton and (non-$b$) jet momenta that appear in this correlation act to some extent as the spin analyzers in the $t$ and $\bar t$ decays. It is well known that the best spin analyzers in the top-quark rest frame are the charged lepton momentum (for semileptonic top decay) and the $d$-quark momentum (for hadronic decay) \cite{Mahlon:1995zn}. Of course, it is not possible to tag the $d$-quark jet in experiment, but at the event generator level we can see how things work. To this effect we define 
$\tilde{A}_{2d,u}$, the counting asymmetry corresponding to $\tilde{\cal O}_{2d,u}=\epsilon(P,p_b+p_{\bar b},p_\ell,p_{d,u})$. These  asymmetries, as the original $A_2$, are interpreted as the sum of processes with $\mu^+, d (~{\rm or}~u)$ from semileptonic decay of $t$ and hadronic decay of $\bar t$, and  processes  with $\mu^-, \bar d  (~{\rm or}~\bar u)$ from semileptonic decay of $\bar t$ and hadronic decay of $t$. 
With the cuts in Eqs.~\ref{cutone}~and~\ref{twocut} and with 
$\tilde{d}= 5 \times 10^{-3}~{\rm GeV}^{-1}$, we find $\tilde{A}_{2d}=1.5\%$ and $\tilde{A}_{2u}=-2.7\%$. 
Interestingly, the asymmetry $\tilde{A}_{2d}$, is {\it smaller} than  $A_2=-3.9\%$, which appears in Table~\ref{t:final1}. To understand what happens, we show in Figure~\ref{f:fig1} the differential distribution of the numerator of $\tilde{A}_2$ with respect to $r$, the ratio of $d$-quark transverse momentum to $u$-quark transverse momentum in $t\to b u \bar{d}$ or $\bar t \to \bar b \bar u d$ decay.
As can be seen in the figure, if one chooses the $d$-quark momentum in the lab frame to construct this particular correlation, there is a partial cancellation between the regions with $p_{Td}>p_{Tu}$ and $p_{Tu}>p_{Td}$. This cancellation is  removed by choosing instead  the hardest jet  resulting in the larger $A_2$. The fact the $A_2$ is larger when using the hardest jet instead of the $d$-quark jet appears to be unique to this correlation.

Using our generated events, we estimate that the $d$-quark jet has a larger $p_T$ than the $u$-quark jet 44\% of the time. 
We can also verify that within statistical errors, the probability of $\bar d$  being the hardest jet in $t\to b u \bar{d}$ is indeed the same as the probability of $d$ being in the hardest jet in $\bar t\to\bar b \bar u d$.

\subsection{Process $p\bar{p}  \to t \bar{t} \to b \bar{b} j_1j_2j_{1'}j_{2'}$}

In this case we only use the cuts of Eq.~\ref{cutone} as there is no missing energy. The results are shown in Table~\ref{t:ubasic2} for about $7.4\times 10^5$ generated events. As expected, the signals $A_1$ and $A_6$ are the same at the parton level and it remains to be seen what dilution there is after hadronization.

\begin{table}[h]
\centering
\begin{tabular}{| c | c | c |c|c|c|c|}
\hline\hline
 & $3/\sqrt{N}$&$A_1$&$A_5$&$A_6$&$A_7$&$A_c$\\
\hline
${\tilde d}$&3.5&-61.2&-54.6&-61.2&38.8&1.1\\
${\tilde f}\sin\phi_f$&3.5&-7.1&-7.8&-7.1&5.8&-1.0\\
${\tilde f}\sin\delta_f$&3.5&-1.8&-1.5&-1.8&-0.5&9.6\\\hline
SM&3.5&0.7&0.5&0.7&1.0&1.1\\
\hline\hline
\end{tabular}
\caption{Integrated asymmetries  for signal 2 with cuts given in Eq.~\ref{cutone} for  $\tilde{d}$,  ${\tilde f}\sin(\phi,\delta)$ $= 5 \times 10^{-3}~{\rm GeV}^{-1}$ in units of $10^{-3}$, and the SM. }
\label{t:ubasic2}
\end{table}

Using our results In Table~\ref{t:ubasic2} for the process $p\bar{p}  \to t \bar{t} \to b \bar{b} j_1j_2j_{1'}j_{2'}$ we find
\begin{eqnarray}
A_1 &=& -0.072\, d_t -0.008\, f_t\sin\phi_f \nonumber \\
A_5 &=& -0.064\, d_t -0.009\, f_t\sin\phi_f \nonumber \\
A_6 &=& -0.072\, d_t -0.008\, f_t\sin\phi_f \nonumber \\
A_7 &=& 0.045\, d_t +0.007\, f_t\sin\phi_f \nonumber \\
A_c &=& 0.011\, f_t\sin\delta_f .
\label{notoprec2}
\end{eqnarray}

\begin{figure}
\centering
\includegraphics[angle=0, width=1.0\textwidth]{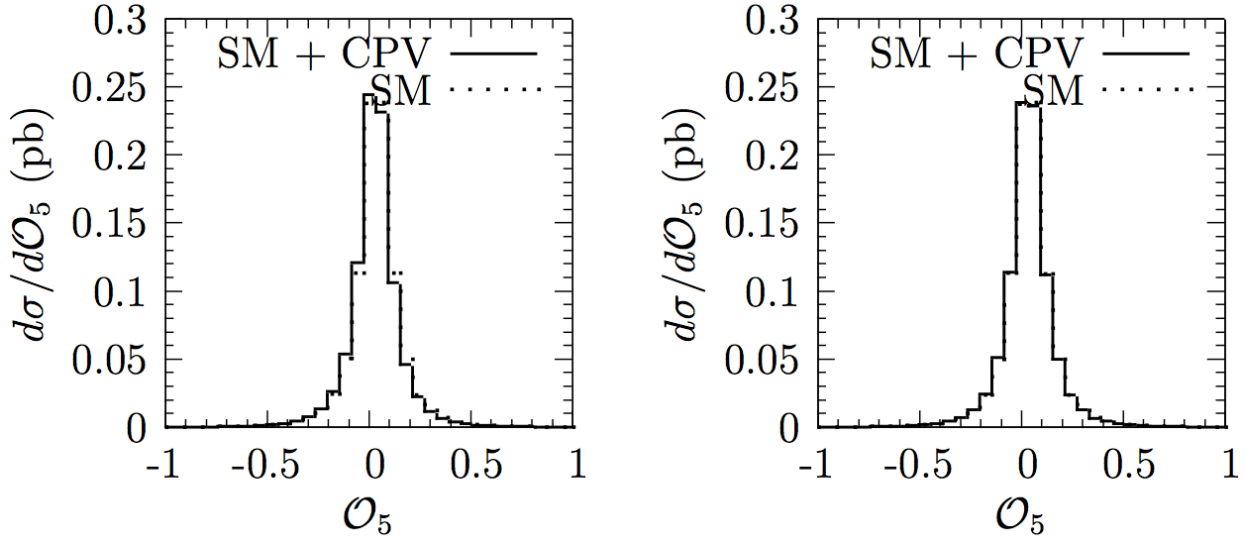}
\caption{Differential distributions $d\sigma/d{\cal O}_5$ for (a) $\tilde{d}=5\times 10^{-3}~{\rm GeV}^{-1}$ and (b) $\tilde{f}\sin\phi_f=5\times 10^{-3}~{\rm GeV}^{-1}$ compared to the SM. ${\cal O}_5$ is given in units of $m_t^4$.}
\label{f:fig3}
\end{figure}

In Figure~\ref{f:fig3} we compare the distributions for $d\sigma/d{\cal O}_5$ induced by (a) $\tilde{d}=5\times 10^{-3}~{\rm GeV}^{-1}$ and (b) $\tilde{f}\sin\phi_f=5\times 10^{-3}~{\rm GeV}^{-1}$ to the SM.

The results in Eq.~\ref{notoprec1} and Eq.~\ref{notoprec2} provide a rough estimate for the sensitivity of the Tevatron to the $CP$ violating anomalous couplings. The existing Tevatron samples of $t \bar t$ events with a double $b$ tag are of the order of 1000 events \cite{tevdata} and this leads to a $3\sigma$ statistical sensitivity to $d_t$ and $f_t$ of order 1. To account for background, we notice that: a) the experimental cuts to select the $t\bar t$ events are the same that will be used for a $CP$ violation study , and b) all the known background processes are $CP$ conserving. The net effect of the background (apart from possible systematic errors that must be studied by the experiments) is to dilute the asymmetries by a factor $(B+S)/S$. The numerator in $A_i$ does not get additional contributions from the background; but the denominator, which counts the total number of events, does. Similarly, the statistical sensitivity decreases by a corresponding factor $\sqrt{(B+S)/S}$. For samples with roughly the same number of background (B) and signal (S) events this amounts to factors of  two.

The gluon fusion initiated dilepton channel at $\sqrt{S}=14$~TeV leads to $A_1= 1.17\,d_t$ \cite{Gupta:2009wu}. Comparing this number to those in Eq.~\ref{notoprec1} and in Eq.~\ref{notoprec2} we see that the dilepton process at LHC is an order of magnitude more sensitive than the lepton plus jets or all hadronic channels at the Tevatron. This is due to two reasons: first the gluon fusion initiated process is more sensitive to the anomalous couplings. For the case of the Tevatron, only about 15\% of top-quark pairs are produced via this mechanism. Numerically we have seen that if we restrict the top-quark pair sample to that originating from gluon fusion, the asymmetries increase roughly by factors of three. The second reason is that the di-lepton channel is  more sensitive to these anomalous couplings.

In addition, the statistical sensitivity of the Tevatron with 1000 events is about five times below that of a $10~{\rm fb}^{-1}$ run at LHC which would produce about 23000 $t\bar t$ dimuon events after the cuts in Eq.~\ref{cutone} and Eq.~\ref{twocut} are applied. Nevertheless, a study with the available Tevatron data would be extremely valuable in understanding the role of systematic errors in measuring $T$-odd asymmetries.

We have performed a series of checks on our numerical analysis as follows. First,  we evaluate the asymmetries for a few values of the anomalous couplings to check that they scale linearly. Second, when the estimated asymmetry is small compared to the $3\sigma$ statistical uncertainty, we repeat the estimate with larger event samples and/or larger values of the anomalous coupling to distinguish between zero asymmetries and numerically small ones. Third, for $t{\bar t}$ pair production at the Tevatron  the parton process with $u\bar{u}$ quarks in the initial state  dominates. We have therefore  estimated the asymmetries using  this parton process only, finding  numbers within 10\%  from the ones obtained when  all initial $q\bar{q}$ and $gg$ states are included. 

\section{Summary and Conclusion}

We have studied the sensitivity of the Tevatron to $CP$ violating anomalous top-quark couplings including its chromo-electric dipole moment $\tilde{d}$. To this effect we have presented a numerical implementation of the results in Ref.~\cite{Antipin:2008zx} using Madgraph for event generation at the parton level. We have considered processes corresponding to $t\bar{t}$ events in the lepton plus jets and all hadronic channels with two b-tags. In order to generate a statistically clean sample we have performed our numerical simulation for a rather large value of the anomalous couplings ($5\times 10^{-3}~{\rm GeV}^{-1}$). Using the fact that all the asymmetries are linear in the anomalous couplings we  present our final results as equations in terms of these couplings, in Eq.~\ref{notoprec1},~\ref{notoprec2}.

Numerically, we find a  $3\sigma$ statistical sensitivity to couplings of order one when normalized to the top-quark mass: $d_t$ and $f_t$, Eq.~\ref{dimcoup}. This sensitivity is about two orders of magnitude below what can be accomplished at the LHC with $10~{\rm fb}^{-1}$. 
These results are based on the assumption that there will be one thousand reconstructed $t\bar{t}$ events. There could be additional inefficiencies in the reconstruction of our specific observables that must be addressed by a careful experimental study. A few comments are in order: models available in the literature to estimate these anomalous couplings typically yield values too small to be observed at the Tevatron; specific models with new sources of $CP$ violation may give contributions to the observables we study that cannot be parametrized by the anomalous couplings. 

With a long term goal of searching for $CP$ violation in $t\bar{t}$ events at the LHC, it is an important exercise to analyze the available Tevatron data and we urge our experimental colleagues to carry out this study. 

\begin{acknowledgments}

This work was supported in part by DOE under contract number DE-FG02-01ER41155. We thank Sehwook Lee and John Hauptman for useful discussions on the D0 $t\bar{t}$ events and David Atwood for useful discussions.

\end{acknowledgments}

\appendix

\section{$T$-odd correlations}

The spin and color averaged matrix element squared that contains the 
$T$-odd correlations can be easily obtained from the results in Ref.~\cite{Antipin:2008zx}. For $CP$ violation in the production process $q\bar{q} \to t \bar{t}$ they can be written as 
\begin{eqnarray}
\left| {\cal M}\right|_{CP}^2  &=&  \, C_1(s,t,u) \, {\cal O}_1 \,+\, C_2(s,t,u) \, {\cal O}_2 \,+\, C_3(s,t,u)\,{\cal O}_3, 
\label{formfactors}
\end{eqnarray}
in terms of the correlations\footnote{Notice that these form factors differ from those defined in Ref.~\cite{Antipin:2008zx} by factors of $t-u$.}
\begin{eqnarray}
{\cal O}_1 &=& \epsilon(p_t,p_{\bar{t}},p_{D},p_{\bar D}) \nonumber \\
{\cal O}_2 &=& \,(t-u) \,\epsilon(p_{D},p_{\bar D},\tilde{p},q) \nonumber \\
{\cal O}_3 &=&\,(t-u) \,\left( \tilde{p} \cdot p_{D} \,
\epsilon(p_{\bar D},p_t,p_{\bar{t}},q)+\tilde{p} \cdot p_{\bar D} 
\,\epsilon(p_{D},p_t,p_{\bar{t}},q) \right).
\label{prodco}
\end{eqnarray}
In Eq.~\ref{formfactors} and in Eq.~\ref{prodco} we have used $s,t,u$, the standard parton level Mandelstam variables for $t\bar t$ production. We have also used the sum and difference of parton momenta
\begin{eqnarray}
\tilde{p}&=&p_1+p_2 \nonumber \\
q&=&p_1-p_2.
\end{eqnarray}

Ref.~\cite{Antipin:2008zx} explicitly gives the result for the case where both $W$s are reconstructed as one jet, in which case the form factors are
\begin{eqnarray}
C^{q{\bar q}}_1(s,t,u)& =& -\frac{16}{9} \, {\tilde d}\,  K_{bb}\, 
m_t\left( \frac{(t-u)^2}{s^2}+4\frac{m_t^2}{s}\right), \nonumber \\
C^{q{\bar q}}_3(s,t,u) &=& -\frac{16}{9} \, {\tilde d}\,  K_{bb}\, 
\frac{m_t}{s^2}, \nonumber \\
C^{q{\bar q}}_2(s,t,u) &=& \frac{s}{2}\, C^{q{\bar q}}_3(s,t,u),
\label{s2cdmqq}
\end{eqnarray}
and $p_D =p_b$, $p_{\bar D} = p_{\bar b}$, 
\begin{equation}
K_{bb} \equiv (\pi^2\alpha_s^2g^4 )\, \left(2-\frac{m_t^2}{M_W^2}\right)^2\, \left(\frac{\pi}{m_t\Gamma_t}\right)^2\, 
\delta(p_t^2-m_t^2)\, \delta(p_{\bar{t}}^2-m_t^2).
\end{equation}
Ref.~\cite{Antipin:2008zx} also indicates how to convert these results into those needed in the case where the $W$s decay leptonically. For  the Tevatron we are interested in two additional cases:

\begin{itemize}

\item Lepton plus jets events. Here one of the $W$ bosons decays leptonically and the other one decays hadronically into two jets. The results for a positively charged lepton, $\ell^+$, follow from the substitutions:
\begin{itemize}
\item In Eq.~\ref{prodco} $p_D \to p_{\ell^+}$ and $p_{\bar D} \to p_{d}$
\item In Eq.~\ref{s2cdmqq} $K_{bb} \to K_{\ell d}$ where 
\begin{eqnarray}
K_{\ell d} &\equiv & 48 \, (\pi^2\alpha_s^2g^8 )\, \left(p_b\cdot p_\nu\right)\left( p_{\bar{b}}\cdot p_{\bar{u}} \right)\, \left(\frac{\pi}{m_t\Gamma_t}\right)^2\left(\frac{\pi}{M_W\Gamma_W}\right)^2 \nonumber \\
&\times&  \delta(p_t^2-m_t^2)\delta(p_{\bar{t}}^2-m_t^2) 
\delta(p_{W^+}^2-M_W^2)  \delta(p_{W^-}^2-M_W^2);
\end{eqnarray}
\end{itemize}
\item For a negatively charged lepton, $\ell^-$, we need the substitutions:
\begin{itemize}
\item In Eq.~\ref{prodco} $p_D \to p_{\bar d}$ and $p_{\bar D} \to p_{\ell^-}$
\item In Eq.~\ref{s2cdmqq} $K_{bb} \to K_{d \ell}$ where 
\begin{eqnarray}
K_{d\ell} &\equiv & 48 \, (\pi^2\alpha_s^2g^8 )\, \left(p_b\cdot p_u\right)\left( p_{\bar{b}}\cdot p_{\bar{\nu}} \right)\, \left(\frac{\pi}{m_t\Gamma_t}\right)^2\left(\frac{\pi}{M_W\Gamma_W}\right)^2 \nonumber \\
&\times&  \delta(p_t^2-m_t^2)\delta(p_{\bar{t}}^2-m_t^2) 
\delta(p_{W^+}^2-M_W^2)  \delta(p_{W^-}^2-M_W^2);
\end{eqnarray}
\end{itemize}

\item For the all hadronic decay the following substitutions are required:
\begin{itemize}
\item In Eq.~\ref{prodco} $p_D \to p_{\bar d}$ and $p_{\bar D} \to p_{d}$
\item In Eq.~\ref{s2cdmqq} $K_{bb} \to K_{dd}$ where 
\begin{eqnarray}
K_{dd} &\equiv & 144 \, (\pi^2\alpha_s^2g^8 )\, \left(p_b\cdot p_u\right)\left( p_{\bar{b}}\cdot p_{\bar{u}} \right)\, \left(\frac{\pi}{m_t\Gamma_t}\right)^2\left(\frac{\pi}{M_W\Gamma_W}\right)^2 \nonumber \\
&\times&  \delta(p_t^2-m_t^2)\delta(p_{\bar{t}}^2-m_t^2) 
\delta(p_{W^+}^2-M_W^2)  \delta(p_{W^-}^2-M_W^2);
\end{eqnarray}
\end{itemize}
\end{itemize}

In our numerical implementation we rewrite all  delta functions as the respective Breit-Wigner distributions behind them, for example:
\begin{eqnarray}
 \left(\frac{\pi}{m_t\Gamma_t}\right) \delta(p_t^2-m_t^2) \to 
 \frac{1}{(p_t^2-m_t^2)^2+\Gamma_t^2m_t^2}.
\end{eqnarray}

When $CP$ violation occurs in the decay vertex, the spin and color averaged matrix element squared containing the $T$-odd correlations was written in Ref.~\cite{Antipin:2008zx} as~\footnote{Note that there is a typo in Ref.~\cite{Antipin:2008zx} where $\phi_f$ and $\delta_f$ are reversed.}
\begin{eqnarray}
|{\cal M}|^2_{T} &=& \, 
\tilde{f}\sin(\phi_f+\delta_f)\, \epsilon(p_t,p_{ b},p_{\ell^+},Q_{t}) +
 \tilde{f}\sin(\phi_f-\delta_f) \,\epsilon(p_{\bar t},p_{ \bar{b}},p_{\ell^-},Q_{\bar{t}}) .
 \label{asymcpdec}
\end{eqnarray} 

The $Q_t$ and $Q_{\bar t}$ that appear in Eq.~\ref{asymcpdec} are linear combinations of available momenta and act as spin analyzers for the $t$ and $\bar{t}$ respectively. For $t\bar{t}$ pairs produced by light $q\bar{q}$ annihilation and when both $W$'s decay leptonically, they are
\begin{eqnarray}
  Q^{q{\bar q}}_{t} &=& K_{\ell \ell} \, \frac{16 m_t}{9s^2} \left\{(4 s m_t^2+(t-u)^2-s^2)p_{\ell^-} +
 2(s p_{\ell^-}\cdot(p_t-p_{\bar t}) -(t-u)p_{\ell^-} \cdot q)p_{\bar t}\right.
 \nonumber \\
& +&\left. 2((t-u)p_{\ell^-}\cdot (p_t+p_{\bar t}) -s p_{\ell^-} \cdot q) q \right\}
\nonumber \\
   Q^{q{\bar q}}_{\bar t} &=&K_{\ell \ell} \, \frac{16 m_t}{9s^2} \left\{ (4sm_t^2+(t-u)^2-s^2)p_{\ell^+} -
 2(s p_{\ell^+}\cdot(p_t-p_{\bar t}) -(t-u)p_{\ell^+} \cdot q)p_{ t}
 \right. \nonumber \\
& -&\left. 2((t-u)p_{\ell^+}\cdot (p_t+p_{\bar t}) +s p_{\ell^+} \cdot q) q
\right\}.
\label{cpdecsqq}
\end{eqnarray}

The necessary replacements to obtain the results relevant for us are:
\begin{itemize}

\item Lepton plus jets events with a positively charged lepton, $\ell^+$:
\begin{itemize}
\item In Eqs.~\ref{asymcpdec},~\ref{cpdecsqq}  $p_{\ell^-} \to p_{d}$
\item In Eq.~\ref{cpdecsqq} $K_{\ell \ell} \to K_{\ell d}$. 
\end{itemize}

\item For a negatively charged lepton, $\ell^-$,
\begin{itemize}
\item In Eq.~\ref{asymcpdec},~\ref{cpdecsqq} $p_{\ell^+} \to p_{\bar d}$
\item In Eq.~\ref{cpdecsqq} $K_{\ell\ell} \to K_{d\ell}$. \end{itemize}

\item For the case in which both $W$'s decay into two jets:
\begin{itemize}
\item In Eq.~\ref{asymcpdec},~\ref{cpdecsqq} $p_{\ell^-} \to p_{d}$, $p_{\ell^+} \to p_{\bar d}$
\item In Eq.~\ref{cpdecsqq} $K_{\ell \ell} \to K_{dd}$. \end{itemize}
\end{itemize}

Corresponding changes are needed for the gluon fusion processes to the results in Ref.~\cite{Antipin:2008zx}.

\end{document}